\documentclass[review,number,sort&compress]{elsarticle}

\usepackage{amsmath}
\usepackage{lineno,hyperref}
\usepackage{here}

\journal{Nuclear Instruments and Methods in Physics Research A }
\biboptions{numbers}

\begin{document}

\begin{frontmatter}

% title
\title{Characterization of the Aging and Excess Noise of a Hamamatsu Fine Mesh Photopentode}

%authors
\author{D. Fujimoto\corref{cor1}}
\ead{fujimoto@phas.ubc.ca}

\author{C. Hearty}
\ead{hearty@physics.ubc.ca}

\cortext[cor1]{Corresponding author.}
\address{University of British Columbia, 6224 Agricultural Road, Vancouver, BC, V6T 1Z1}

% ABSTRACT
\begin{abstract}
The excess noise factor and the aging characteristics of 16 Hamamatsu R11283 photopentodes have been tested. These fine-mesh phototubes are to be paired with pure CsI scintillation crystals considered for use in the endcap calorimeter of the Belle~II detector. The average excess noise factor was found to be $1.9\pm0.1\pm0.4$. The electronic noise of a custom preamplifier produced by the University of Montreal was found as a consequence of this measurement and was $1730\pm33$ electrons, in agreement with previous values.  On average, the gain~$\times $~quantum efficiency was reduced to $92\pm3$~\% of the initial value after passing an average of $7$~C through the anode. This corresponds to 70 years of standard Belle~II operation.
\end{abstract}

%keywords

\end{frontmatter}

% INTRODUCTION
\section{Introduction}
\begin{figure}[t]
    \centering
    \includegraphics[keepaspectratio=true,width=9cm]{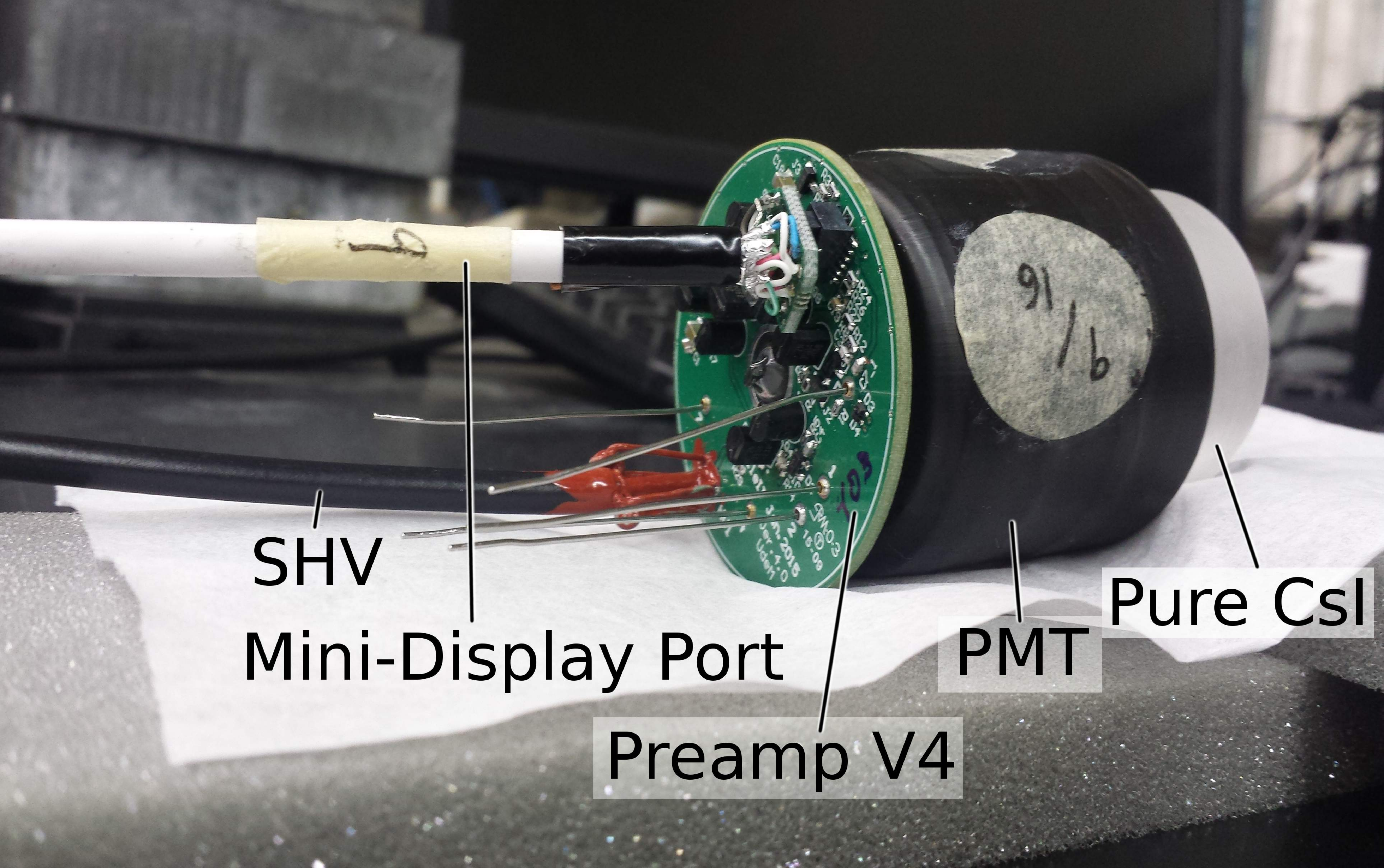}
    \caption{R11283 PMT with version 4 preamp produced by the University of Montreal~\cite{preamp} and glued pure CsI puck (AMCRYS) for the aging study. The preamp functions and output are powered and transmitted by a mini-display port cable, whereas the voltage divider has a SHV cable soldered to it. The preamp has a diameter of 5~cm and a length of about 3.8~cm, excluding the flying leads.}
    \label{PP_CsI_preamp}
\end{figure}

One of the upgrade considerations for the Belle~II detector, situated at KEK in Tsukuba, Japan, is to replace the current thallium-doped cesium iodide (CsI(Tl)) scintillation crystals in the endcap electromagnetic calorimeter (ECL) with pure CsI~\cite{belle2Tech}. A primary goal of this exchange is to reduce pileup due to the increased luminosity of the SuperKEKB accelerator. While pure CsI has a shorter scintillation time constant, it also has a reduced light yield, with the emission spectrum peaking in the UV range rather than in the visible range~\cite{csiPure,csiDoped}. Therefore, the new crystals will need new photosensors. Under consideration for the new photosensor is the R11283 photomultiplier tube (PMT), developed by Hamamatsu Photonics for this project. This model has five flying leads and for this reason is often referred to as a photopentode. The PMT is a head-on type with three fine mesh dynodes, UV transparent window, and a bialkali photocathode; similar to previously tested PMTs for CsI scintillation crystals, although this model is of much lower gain and has fewer dynodes \cite{PP_performance}. The photocathode of the R11283 has a minimum effective diameter of $39$~mm, and a wavelength of maximum response of $420$~nm~\cite{R11283Tech}. Pure CsI has an emission maximum at $315$~nm. The PMT is shown in Fig.3~\ref{PP_CsI_preamp}. 

%PP photo

Magnetic fields decrease the performance due to changes in the inter-dynode electron path. These fine mesh PMTs will operate in the 1.5~T Belle~II axial magnetic field, reducing the nominal gain by a factor of 3.5~\cite{Kuzmin_ECL}. The average nominal gain of the 16 PMTs at an operating voltage of $-1000$~V was $255\pm11$. As summarized in Table \ref{hamaTbl}, Hamamatsu provides a variety of measurements at $-750$~V with purchase. 

%hamamatsu summarized quantities
\begin{table}[t]
\centering
\caption[Hamamatsu Measured Quantities]{Average Hamamatsu quantities which were measured at $-750$~V and the PMT internal gain for a population of 16 PMTs.}
\begin{tabular}{lll}
\hline
Quantity & Average & Range \\ \hline
Anode Luminous Sensitivity ($\mu$A/lm)		& $14000\pm1000$ 		& $[8750.0,26100.0]$\\
Cathode Luminous Sensitivity ($\mu$A/lm) 	& $83\pm3$		 		& $[69.3,106.0]$\\
Cathode Blue Sensitivity Index				& $9.4\pm0.1$			& $[8.70,10.10]$\\
Dark Current (nA) 							& $0.011\pm0.007$		& $[0.00,0.03]$\\
Gain at $-750$~V 							& $170\pm7$				& $[125,276]$\\
Gain at $-1000$~V 		     				& $255\pm11$			& $[183,367]$\\\hline
\end{tabular}
\label{hamaTbl}
\end{table}

\begin{figure}[ht]
    \centering
    \includegraphics[keepaspectratio=true,width=9cm,trim= 1cm 0cm 1cm 5cm, clip=true]{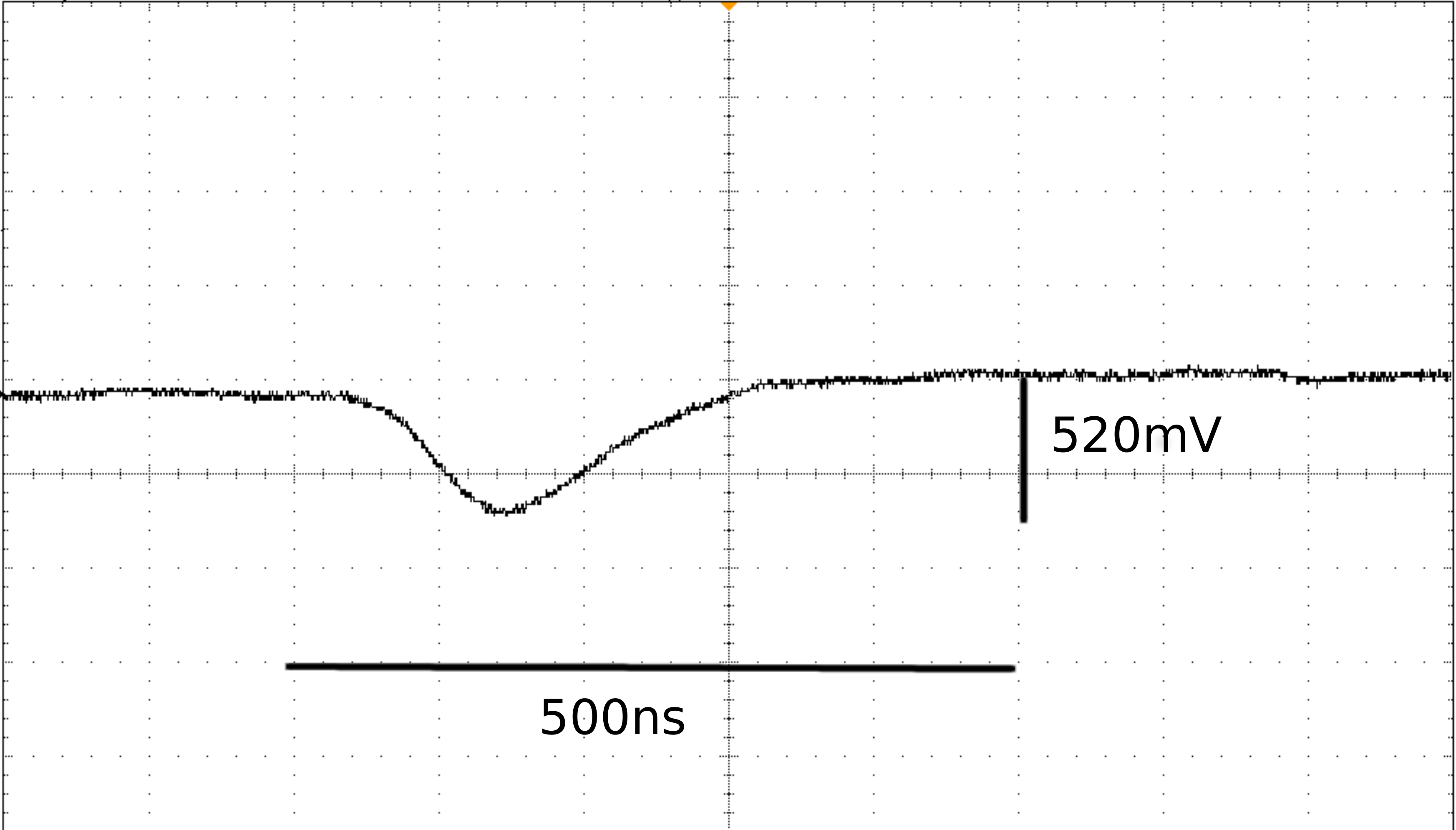}
   \caption{Typical single shot waveform produced by the University of Montreal shaper from a signal produced in pure CsI. The signal has a slightly longer peaking time due to the time constant of the scintillation material.}
    \label{screenshot}
\end{figure}

The PMT readout electronics were designed and produced by the University of Montreal~\cite{preamp} and consists of a preamp (version 4) and a shaper for every PMT. The preamp has a gain of $0.5$~V/pC and the shaper has a shaping time of 50~ns. For a step-function input, the signal produced by the shaper has a peaking time of $200$~ns. The combination of the two components produces a signal whose amplitude is proportional to the charge at the anode (Fig.~\ref{screenshot}). The board connected directly to the PMT (Fig.~\ref{PP_CsI_preamp}) houses both the voltage divider to power the PMT dynodes, and the preamp electronics. The shapers house and are powered by a motherboard which also provides the correct voltages to the preamp electronics. 

Additional details on the measurements presented can be found in Ref.~\cite{thesis}.

% METHODS
\section{Methods}
\begin{figure}[ht]
    \centering
    \includegraphics[keepaspectratio=true,width=9cm,trim= 0cm 0cm 0cm 0cm, clip=true]{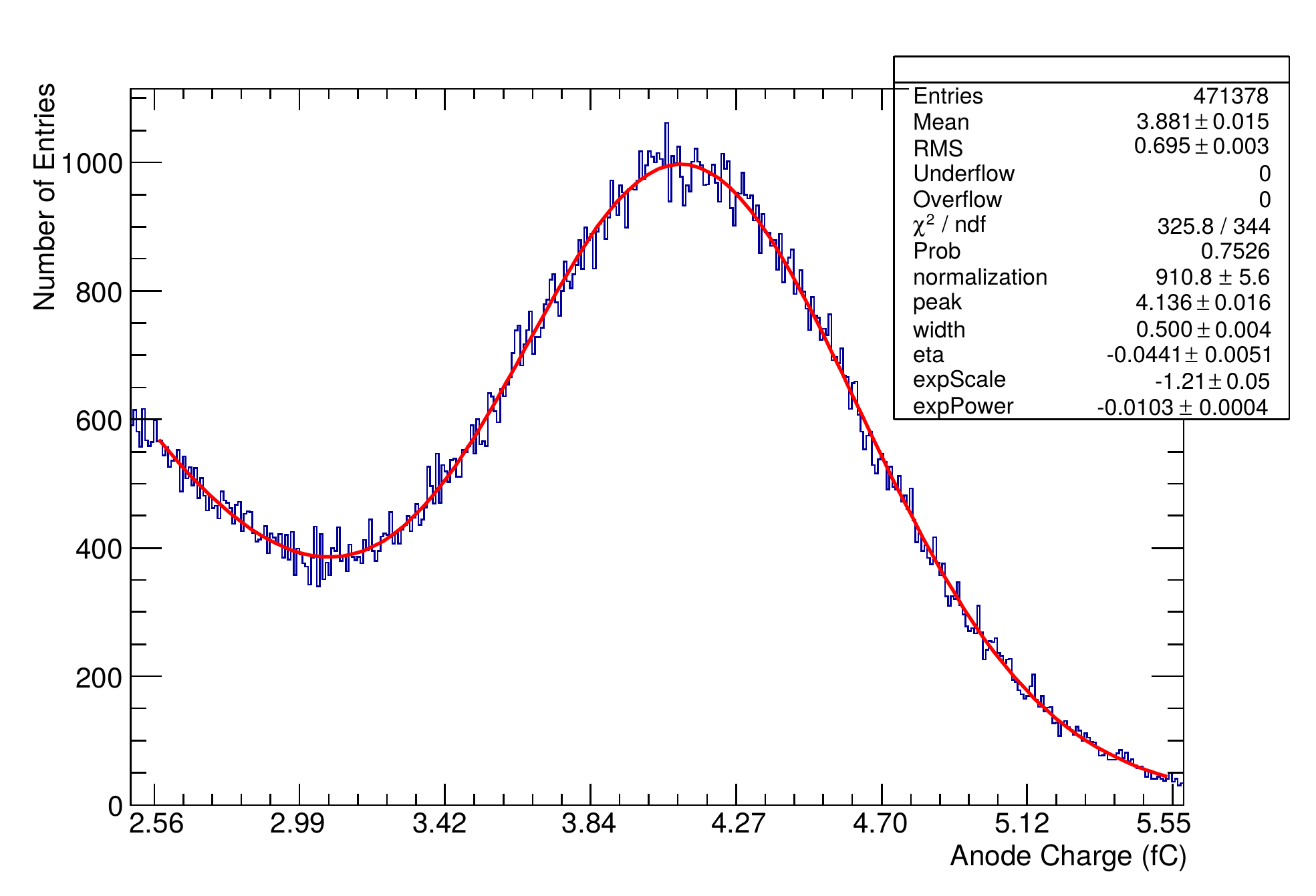}
   \caption{Histogram of peak-sensing ADC output with Novosibirsk + exponential fit for the spectrum produced by gammas from a $^{137}$Cs source. A large background peak for anode charge less than $2.5$~fC has been omitted for clarity.}
    \label{novoFit}
\end{figure}

The shaper output was fed into a peak-sensing ADC (LeCroy L2259B) and the output was histogrammed using the MIDAS program~\cite{MIDAS}. The histogram was then fitted with the sum of an exponential and a Novosibirsk function. The exponential roughly describes the background, which was primarily due to backscatter. The Novosibirsk function is an asymmetric Gaussian-like function with four parameters: height, peak location, width, and an asymmetry parameter~\cite{novosibirsk}. Fig.~\ref{novoFit} shows an example of fitting this sum to the spectrum produced by the $662$~keV decay of $^{137}$Cs.

%screenshot

% Novo Fit

The response of the system to energy deposits in pure CsI is linear. Using several calibration sources, the value of the peak location at zero energy deposited was extrapolated and attributed to a DC offset in the peak-sensing ADC. This pedestal was then subtracted from the measurements prior to any further manipulation. The charge at the anode is proportional to the peak location, and the uncertainty to the width of this distribution. To calibrate the charge at the anode, the calibration test pulse feature of the preamp was used. This allows for a known amount of charge to be injected into the preamp, which is then processed as every other signal. From this calibration, the relationship between the ADC binning and the charge at the anode was found. 

% EXCESS NOISE FACTOR 
\section{Excess Noise Factor}
The excess noise factor is a common index for estimating the performance of photosensors~\cite{hamamatsuHandbook}. This factor describes the uncertainty introduced into the system as a result of the electron multiplication process: 
\begin{equation}
\left(\frac{\sigma_a}{N_a}\right)^2 = F \cdot \left(\frac{\sigma_c}{N_c}\right)^2,  \label{eqnF}
\end{equation}
where the subscripts $c$ and $a$ denote the cathode and anode respectively. Measured at location $x$, $N_x$ and $\sigma_x$ are the number of electrons and the signal width respectively, as determined from the Novosibrisk fit. $F$ is the excess noise factor. Typically, the excess noise factor is larger for fine mesh PMTs than standard PMTs. Recognizing the Poisson nature of the photoelectrons and that the measured width at the anode contains contributions from both the constant electronic noise and the excess noise factor, Equation \ref{eqnF} can be written as: 
\begin{equation}
\sigma^2_m = F\cdot N_c + \sigma_o^2, \label{eqnF_lin}
\end{equation}
where the internal PMT gain has been applied to put all relevant quantities in units of number of electrons at the photocathode (photoelectrons). Here, $\sigma_m$ is the measured width at the anode and $\sigma_o$ is the contribution of the electronic noise. From this, the excess noise factor can be easily found by varying the light intensity ($N_c$) and measuring the resulting distribution width ($\sigma_m$).

%setup
\begin{figure}[ht]
    \centering
    \includegraphics[keepaspectratio=true,width=9cm,trim= 2cm 4cm 3cm 3cm, clip=true]{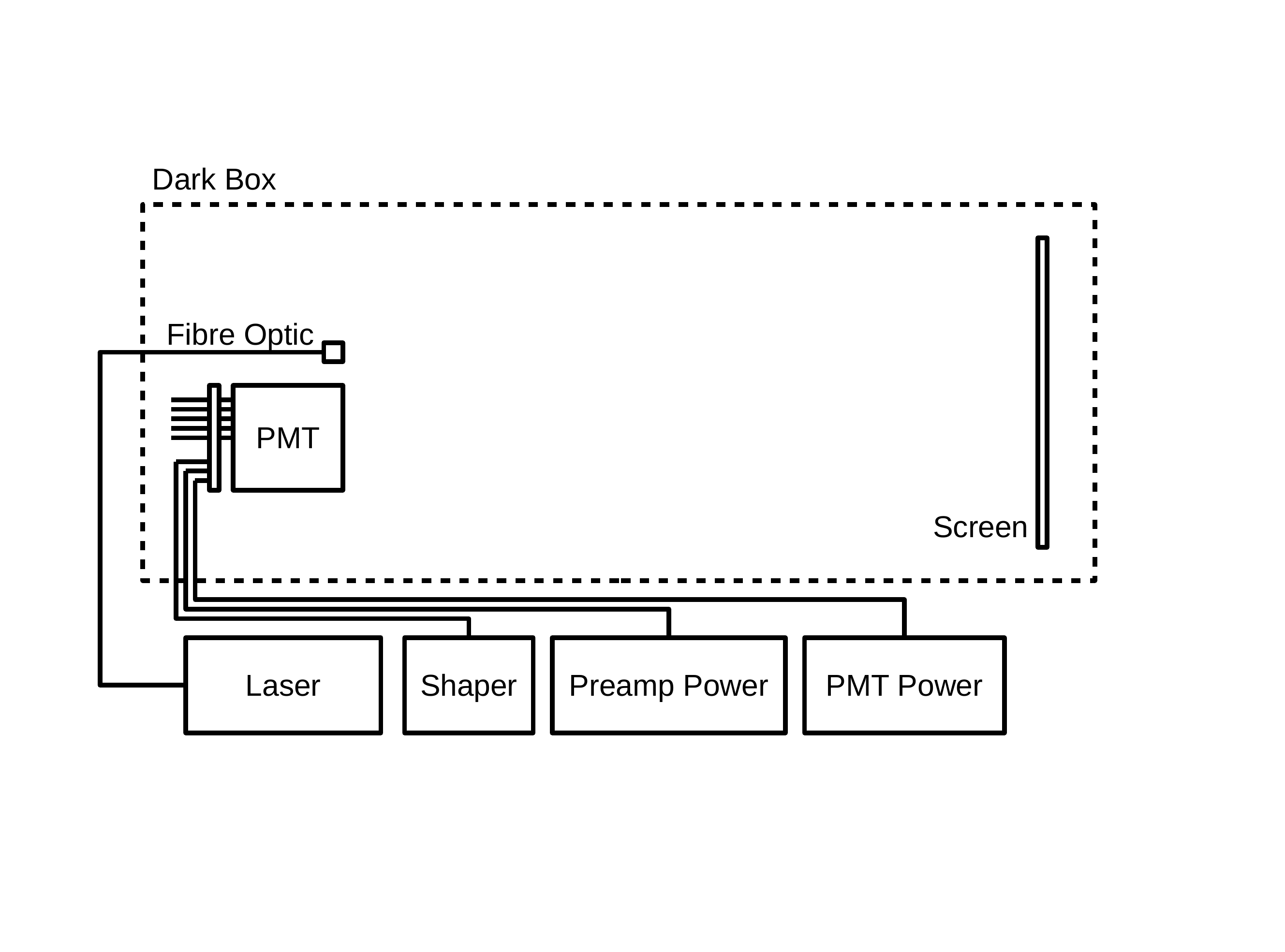}
   \caption{Setup for the excess noise measurement. A UV laser was reflected off of a screen to provide uniform incident light of controlled intensity.}
    \label{Fsetup}
\end{figure}

%analysis
\begin{figure}[ht]
    \centering
    \includegraphics[keepaspectratio=true,width=9cm,trim= 0cm 0cm 0cm 1cm, clip=true]{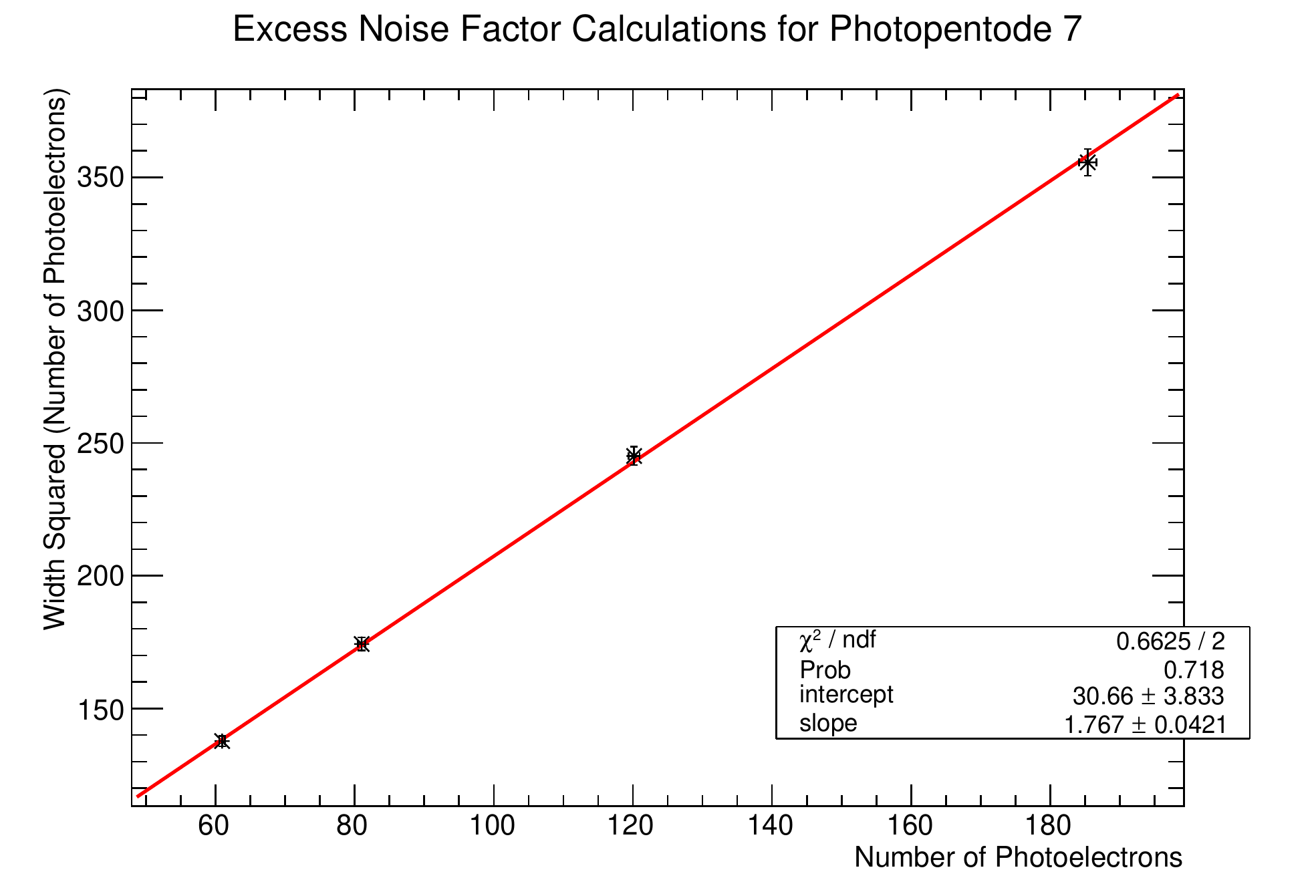}
   \caption{The excess noise factor is given by the slope of the fitted linear line as seen in Eq.~\ref{eqnF_lin}. The intercept gives the electronic noise in units of number of photoelectrons.}
    \label{Fanalysis}
\end{figure}

To this end, a $405$~nm laser was pulsed at $350$~Hz to illuminate the PMT (Fig.~\ref{Fsetup}). The laser was first reflected off of a diffusive white screen to provide uniform light. The light intensity was controlled via the laser voltage and the screen-to-PMT distance. Using a single preamp, the excess noise factor was found for all 16 PMTs.

Fig.~\ref{Fanalysis} shows the results of this analysis for one of the PMTs. The slope is the excess noise factor, and the intercept is the electronic noise in units of photoelectrons. The operating voltage for these measurements was $-1000~$V. On average, the excess noise factor was found to be $1.9\pm0.1\pm0.4$, where the statistical error of $0.1$ is the standard deviation across the 16 PMTs. The systematic error of $0.4$ is due to the $25\%$ uncertainty in the value of the capacitor used in the test pulse calibration~\cite{preamp}. The average electronic noise was $1730\pm33$ electrons at the anode, which is in good agreement with previous measurements~\cite{preamp}. The range of the excess noise factors was $1.8$~--~$2.1$, whereas the range of the electronic noise was $1526$~--~$1913$ electrons at the anode. Given that the light yield of the CsI crystal is $85$ photoelectrons per MeV deposited~\cite{csiYield}, this electronic noise corresponds to an equivalent noise energy of about $80$~keV.

To estimate the impact of magnetic fields on the excess noise, the PMT gain was reduced by lowering the operating voltage. Above a gain of 55, the PMT was seen to have a constant excess noise factor. Below this, the excess noise factor increases non-linearly with decreasing gain, rising to $3.5\pm0.1\pm0.4$ at a gain of 25. In comparison, two avalanche photodiodes (APD) from the Hamamatsu S8664 series have been measured to have an excess noise factors of 3.4 and 5.1~\cite{APD15}. These APDs are also being studied for pure CsI application as a competing option for the Belle~II endcap ECL upgrade. 

% AGING AND LIFETIME
\section{Aging}

Of importance to the Belle~II experiment is the effects of the PMT aging. Given that the Belle~II endcap ECL will be in an axial magnetic field of approximately $1.5$~T, and that the PMTs will be within $12.4^\circ$--$31.4^\circ$ to this field \cite{belleTech}, it is expected that the gain of the PMTs will drop by about a third~\cite{Kuzmin_ECL}. To simulate this, the aging process was performed with the operating voltage of the PMTs set to $-491$~V, reducing the average gain to $85\pm3$ or one third of the nominal gain at $-1000~$V. 

\begin{figure}[h]
    \centering
    \includegraphics[keepaspectratio=true,width=10cm,trim= 1cm 1cm 1cm 1cm, clip=true]{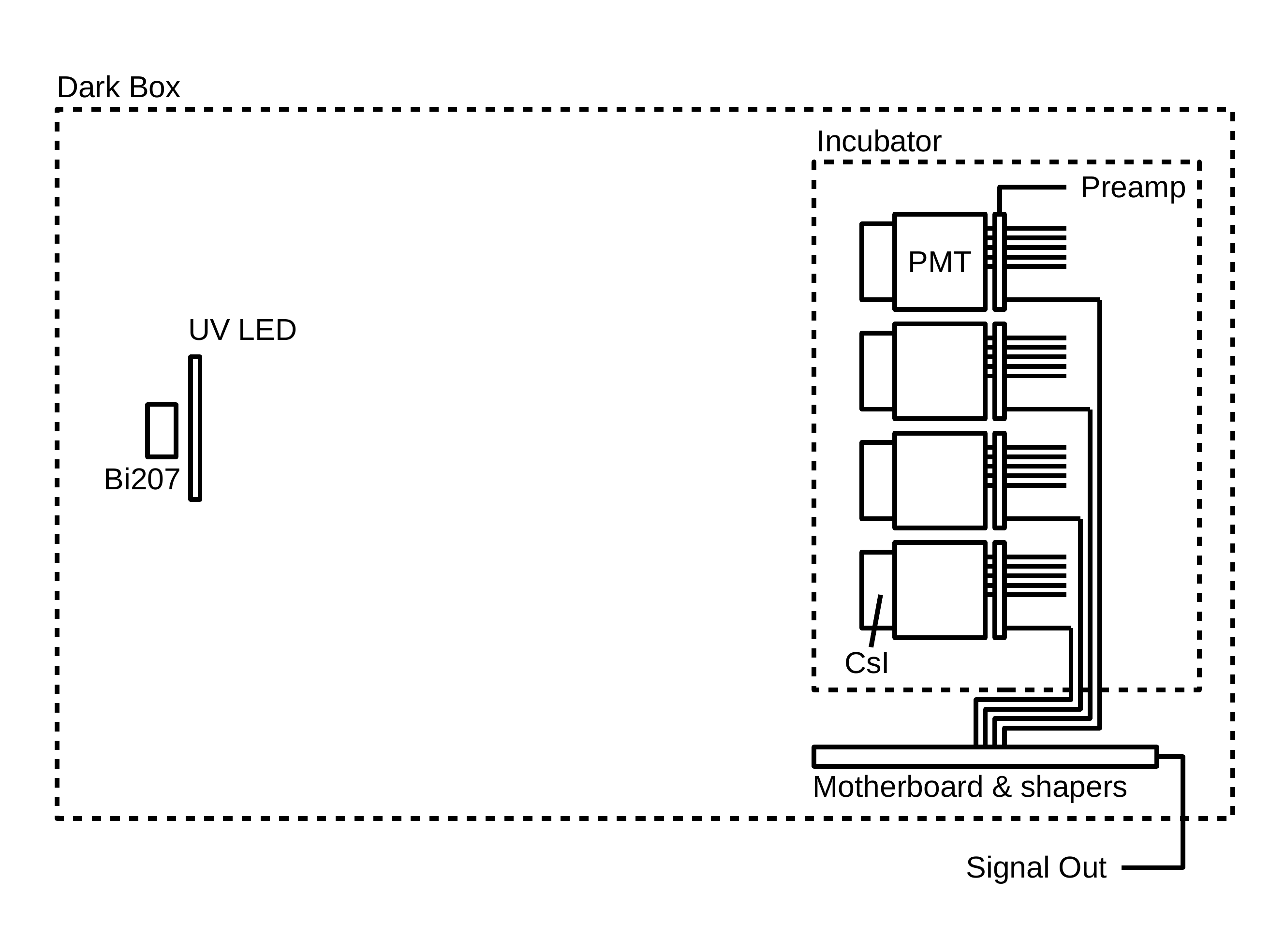}
   \caption{The experimental setup for the aging measurements. A 4$\times$4 array of PMTs was encased in an incubator, with the temperature and humidity controlled. Aging was accelerated with a UV LED, and a $^{207}$Bi source was used to track the PMT performance. }
    \label{AgingSetup}
\end{figure}

The performance of the PMT was characterized by the gain$\times$quantum efficiency, which was found from the slope of the peak ADC bin as a function of the energy deposited in the CsI. The change in the gain$\times$quantum efficiency relative to the initial value was monitored as a function of the charge passed through the anode, and also as a function of the real lab time elapsed. Light from a UV LED (335~nm) was used to age the PMTs, which were arrayed in a 4$\times$4 array (Fig.~\ref{AgingSetup}). The PMTs were encased in an incubator with a UV transparent acrylic window to maintain the temperature at $37\pm2~^\circ$C. The relative humidity was kept within $15$~--~$20$~\% by means of desiccant and a slow influx of N$_2$ gas into the incubator. One of the 16 PMTs was capped with black rubber to prevent aging and act as a control. This PMT was also used to correct for the residual variation in temperature. The relative peak location response with temperature was roughly linear and varied at a rate of $-1.2\pm0.2~\%/^\circ$C. Probably due to the poor thermal contact between dynodes and the environment exterior to the PMT, there is about a $8$~--~$10$~h delay before the PMTs reached thermal equilibrium.

%setup

To track the performance, a constant light source was produced by gluing pure CsI pucks to the faces of the PMTs (Fig.~\ref{PP_CsI_preamp}) and triggering scintillation light with a $^{207}$Bi source. The glue used was TSE3032 silicone rubber produced by Momentive Performance Materials, which has an index of refraction of $1.406$ for $589$~nm light~\cite{TSE3032}. In comparison, the index of refraction of pure CsI is $1.95$ at the emission maximum of $315$~nm~\cite{csiPure}. The CsI cylindrical pucks were manufactured by AMCRYS and were approximately the same diameter as the photocathode. 

The measurement of the current was enabled by a modified preamp from the University of Montreal. This preamp did not perform any of the signal processing of the regular preamps, but rather contained a resistor such that the current could be measured with a Keithley 6485 Picoammeter. Prior to aging, a current baseline was established for each PMT using constant incident light. Using the baseline, only one PMT was needed to track the current through the anode and the current could be estimated for the other PMTs. 

\begin{figure}[H]
    \centering
    \includegraphics[keepaspectratio=true,width=17cm,trim= 0cm 0cm 0cm 0cm, clip=true, angle=90]{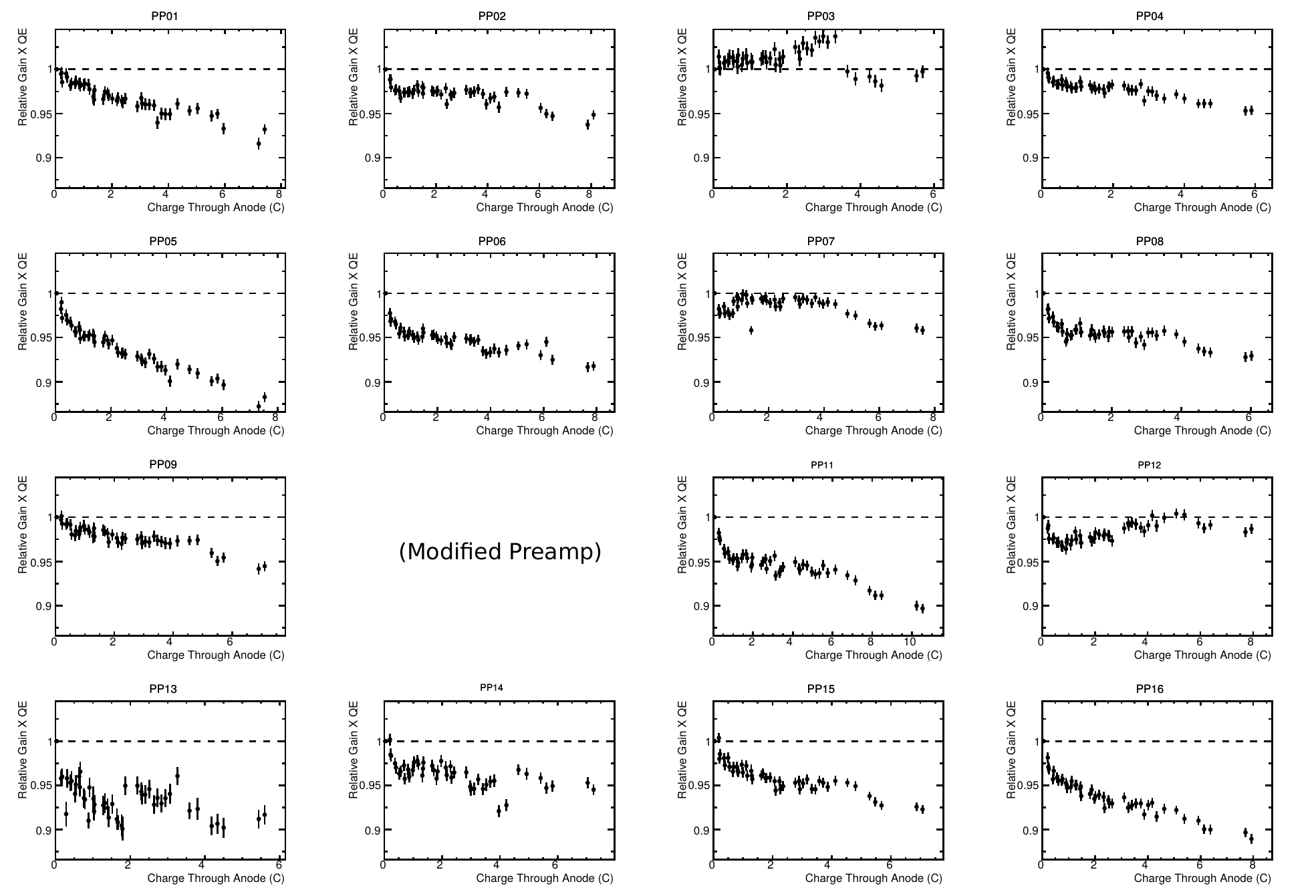}
   \caption{PMT gain$\times$quantum efficiency relative to the first measurement, as a function of the charge passed through the anode, after temperature corrections. The horizontal dotted line marks the initial relative value of one. PP03 is the capped control PMT. }
    \label{gSvQ_All}
\end{figure}

The $^{207}$Bi source was chosen for its two easily visible decays at $0.570$~MeV and $1.064~$MeV. Given the linear response of the PMT with energy, these two peaks were used to establish the gain$\times$quantum efficiency, which can be seen as a function of the integrated current in Fig.~\ref{gSvQ_All}. 

In Fig.~\ref{gSvQ_All}, the PMT in the third column of the first row (PP03) is the control PMT and was not aged. There are a few different observed behaviors in the PMT aging. Some exhibited a burn-in period where the performance decreased rapidly during the first coulomb of charge, then appeared to stabilize. Some appeared to age continuously throughout, whereas others exhibited little to no aging or even experienced an improvement in the performance. The control PMT did not see any significant change in performance at the end of the aging process. 

%analysis

%analysis
\begin{figure}[H]
    \centering
    \includegraphics[keepaspectratio=true,width=17cm,trim= 0cm 0cm 0cm 0cm, clip=true, angle=90]{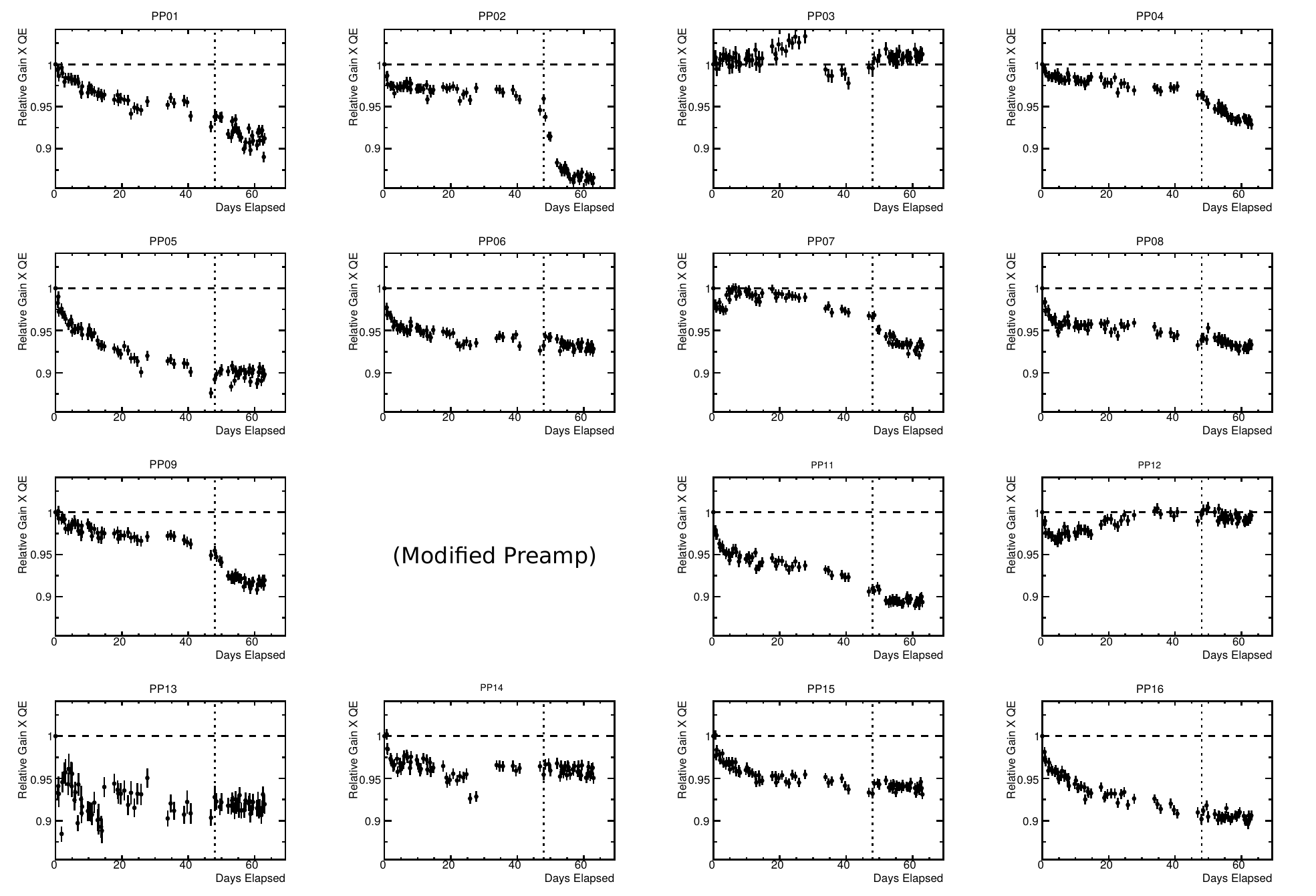}
   \caption{PMT gain$\times$quantum efficiency relative to the first measurement, as a function of lab time elapsed, after temperature corrections. The vertical dotted line marks the point where the UV LED was turned off and aging was halted. The horizontal dotted line marks the initial relative value of one. }
    \label{gSvT_All}
\end{figure}

After 48 days of aging, the LED was turned off and the stability monitored (Fig.~\ref{gSvT_All}). Even after aging, some of the PMTs continued to see a decrease in performance. Of note, one of the 14 aged PMTs (PP02) saw nearly 10~\% decrease in performance within a week. This degradation is worrying as it is rapid enough to be difficult to track via a physics-based calibration system. The control PMT did not observe any significant change in performance after the end of the aging process. At the end of the aging process an average of $7.4\pm1.2$~C was passed through the PMT anodes and the average PMT performance was reduced to $92\pm3$~\% of the initial value. 

\begin{figure}[H]
    \centering
    \includegraphics[keepaspectratio=true,width=\textwidth,trim= 0cm 0cm 0cm 0cm, clip=true]{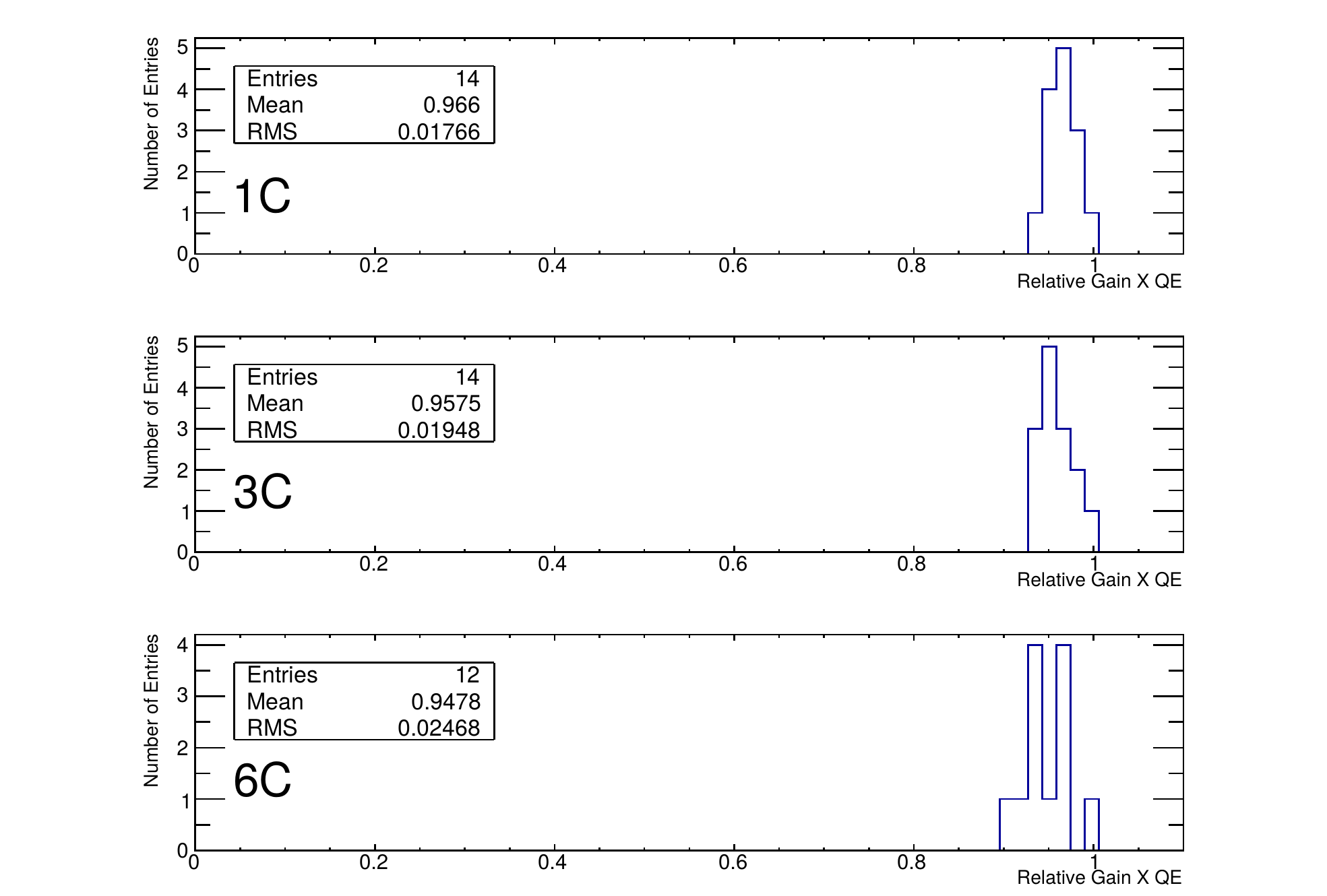}
	\vspace{-1cm}
	\caption{Histograms of the PMT gain$\times$quantum efficiency relative to the first measurement, at various stages in the aging process. The charge given is the anode charge after temperature corrections. }
    \label{performanceHist}
\end{figure}

By definition, each PMT has a relative performance of 1 at zero charge. Other than at this point, the PMTs appear to be scattered normally about the mean (Fig.~\ref{performanceHist}). Due to the internal gain of the PMTs, an equal amount of incident light does not produce the same number of electrons at the anode, and as a result there are fewer statistics available at the larger anode charges.  

%analysis - histo

\begin{figure}[H]
    \centering
    \includegraphics[keepaspectratio=true,width=\textwidth,trim= 0cm 0cm 0cm 0cm, clip=true]{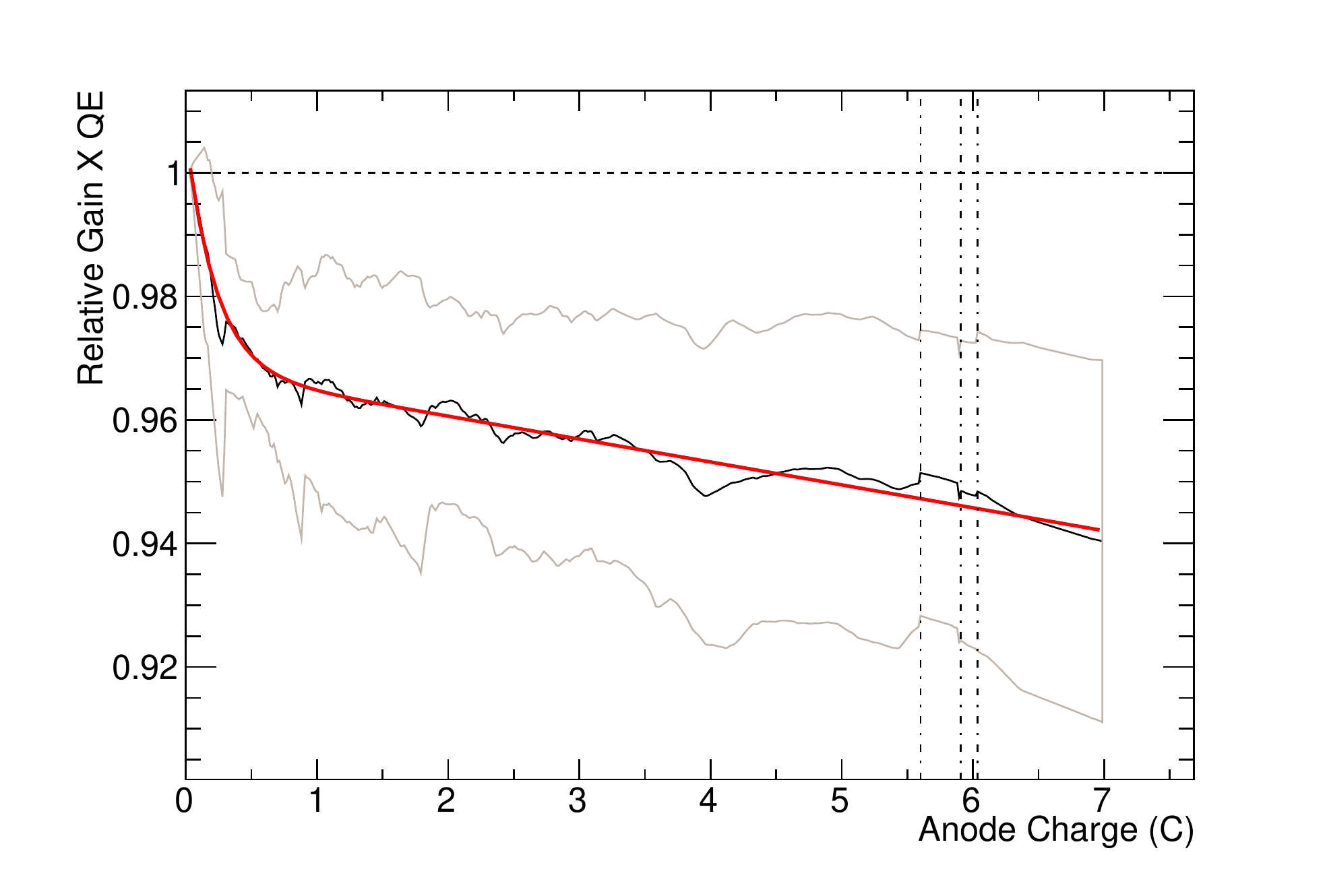}
	\vspace{-1cm}
	\caption{Average relative PMT gain$\times$quantum efficiency as a function of anode charge up to $7$~C, with RMS envelope and functional overlay. The charge given is after temperature corrections, and the function is that of Eq.~\ref{fit}. The vertical lines denote where the number of PMTs present in the average drops by one (starting with 14). This occurs at $5.6$~C, $5.9$~C, and $6.0$~C. Beyond $7$~C the number of dropped PMTs increases rapidly.}
    \label{fracVCharge}
\end{figure}

Fig.~\ref{fracVCharge} shows the average gain$\times$quantum efficiency as a function of charge, with an envelope denoting the RMS. These values were found by taking a linear interpolation to account for the gaps in the measurement. As mentioned earlier, not all of the PMTs saw an equal amount of charge. At $5.6$~C, the total cumulative charge passed through the anode of the lowest gain PMT, the relative gain$\times$quantum efficiency was $95\pm2$~\% across all 14 PMTs. As was seen in Fig.~\ref{performanceHist}, the RMS grows steadily with anode charge. 

Empirically, the sum of a linear and exponential function was chosen to be fitted to the curve in Fig.~\ref{fracVCharge} in the range of $[0,5.5]$~C, using the error in the mean as the fit weighting. The resulting function is given by 
\begin{equation} \label{fit}
[\mbox{Rel. Gain}\times\mbox{QE}] = 0.968 - 0.0037q + 0.0383e^{-q/0.23}.
\end{equation}
This shows that there exists a burn-in period, modelled by the exponential term, after which the PMT experiences a linear decay in performance of $-0.4$~\%/C. The burn-in period lasts for $1.05$~C, after which the exponential factor scales its coefficient by less than $1$~\%. By $1.57$~C the exponential contributes less than $0.1$~\% to the function. 

%analysis - mean vs Q

% SUMMARY
\section{Summary}
The initial quantities measured by Hamamatsu prior to shipping can be found in Table \ref{hamaTbl}. Using processing electronics developed by the University of Montreal, the excess noise factor and aging properties of the R11283 photopentode were studied. The average excess noise factor for 16 PMTs was found to be $1.9\pm0.1\pm0.4$ and the average electronic noise was measured to be $1730\pm33$ electrons at the anode. This electronic noise corresponds to an equivalent noise energy of 80~keV. The phototube aging was also studied by passing a large amount of charge through the anode, by exposing the photocathode to a large amount of incident light. At a reduced gain of $85\pm3$, 14 PMTs were aged with an average of $7.4\pm1.2$~C passed through the anode, reducing the average performance of the PMTs to $92\pm3$~\% of the initial measurement. Of the 14 aged PMTs, only one showed signs of rapid aging that could be a problem for some calibration systems. The average change in performance is characterized by an exponential burn-in period that lasts approximately $1.05$~C, after which the performance degrades linearly by $-0.4$~\%/C.

% ACKNOWLEDGEMENTS 
\section*{Acknowledgments}
This work was supported by the technical support staff at TRIUMF, in particular P. Amaudruz who developed the pulsed UV laser and aided in the setup of MIDAS. Additionally, J.P. Martin, N.A. Starinski, and P. Taras of the University of Montreal designed and produced the preamp, motherboard, and shaper electronics. D. Jow aided in the gluing of the CsI to the PMTs. Funding for this work was provided by NSERC. 

% REFERENCES
\section*{References}
\bibliography{fineMeshPP}

\end{document}